\documentclass[11pt,prb,showpacs,amsmath,amssymb,superscriptaddress]{revtex4-1}
\usepackage{graphicx}
\usepackage{hyperref}
\usepackage{xcolor}
\usepackage{color}
\makeatother
\usepackage{multirow}

\begin{document}

\title{Crystal Hall effect in collinear antiferromagnets}

\author{Libor \v{S}mejkal}
\affiliation{Institut f\"ur Physik, Johannes Gutenberg Universit\"at Mainz, D-55099 Mainz, Germany}
\affiliation{Institute of Physics, Czech Academy of Sciences, Cukrovarnick\'{a} 10, 162 00 Praha 6, Czech Republic}
\affiliation{Faculty of Mathematics and Physics, Charles University in Prague, Ke Karlovu 3, 121 16 Prague 2, Czech Republic}
\author{Rafael Gonz\'{a}lez-Hern\'{a}ndez }
\affiliation{Grupo de Investigaci\'{o}n en F\'{\i}sica Aplicada, Departamento de F\'{i}sica, Universidad del Norte, Barranquilla, Colombia}
\affiliation{Institut f\"ur Physik, Johannes Gutenberg Universit\"at Mainz, D-55099 Mainz, Germany}
\author{Tom\'{a}\v{s} Jungwirth}
\affiliation{Institute of Physics, Czech Academy of Sciences, Cukrovarnick\'{a} 10, 162 00 Praha 6, Czech Republic}
\affiliation{School of Physics and Astronomy, University of Nottingham, Nottingham NG7 2RD, United Kingdom}
\author{Jairo Sinova}
\affiliation{Institut f\"ur Physik, Johannes Gutenberg Universit\"at Mainz, D-55099 Mainz, Germany}
\affiliation{Institute of Physics, Czech Academy of Sciences, Cukrovarnick\'{a} 10, 162 00 Praha 6, Czech Republic}

\date{\today}

\begin{abstract}
\textbf{Electrons, commonly moving along the applied electric field, acquire in certain magnets a dissipationless transverse velocity. 
This spontaneous Hall effect, discovered more than a century ago,  has been understood in terms of the time-reversal symmetry breaking by the internal spin-structure of a ferromagnetic, noncolinear antiferromagnetic or skyrmionic form. 
Here we identify previously overlooked robust Hall effect mechanism arising from collinear antiferromagnetism combined with nonmagnetic atoms at non-centrosymmetric positions. 
We predict a large magnitude of this crystal Hall effect in a room-temperature collinear antiferromagnet RuO$_2$ and catalogue, based on our symmetry rules, extensive families of material candidates.
We show that the crystal Hall effect is accompanied by the possibility to control its sign by the crystal chirality.
We illustrate that accounting for the full magnetization density distribution instead of the simplified spin-structure sheds new light on symmetry breaking phenomena in complex magnets and opens an alternative avenue towards quantum materials engineering for low-dissipation nanoelectronics.}
\end{abstract}

\maketitle

\section*{Introduction} 
The spontaneous Hall voltage arises when the electrons gain transverse velocity due to the certain internal magnetic structures.
The associated Hall conductivity is the antisymetric dissipationless part of the conducitivity tensor, which corresponds to the Hall pseudovector $\boldsymbol\sigma$, that determines the Hall current\cite{Hall1879,Nagaosa2010,Shtrikman1965}:
\begin{equation}
\textbf{j}_{H}=\boldsymbol\sigma\times \textbf{E}.
\label{pseudovector}
\end{equation}
Here $\textbf{E}$ is the applied electric field, $\boldsymbol\sigma=(\sigma_{zy},\sigma_{xz},\sigma_{yx})$, $\sigma_{ij}$ are the antisymmetric Hall conductivity components, and $\textbf{j}_{\rm H}$ is the Hall current transverse to $\textbf{E}$ and $\boldsymbol\sigma$. 
Apart from being odd under time-reversal ($\mathcal{T}$), Eq.~(\ref{pseudovector}) explicitly highlights that the Hall effect transforms like a pseudovector under spatial symmetry operations, i.e., it transforms like a magnetic dipole moment. 
This implies that the \textit{spontanous} Hall effect (in the absence of an external field)  can occur only in materials with a magnetic space group (MSG), in which a net magnetic moment is allowed by symmetry\cite{Kleiner1966,Nagaosa2010,Seemann2015,Suzuki2017,Grimmer1993}. 
In fact, since $\boldsymbol\sigma$ is invariant under the spatial inversion ($\mathcal{P}$), its components allowed by symmetry can be determined from the magnetic Laue group (MLG) \cite{Kleiner1966,Nagaosa2010,Seemann2015}.
 
In the conventional picture of (spontaneous) anomalous Hall effect  (AHE) in ferromagnets, the Hall voltage arises due to the asymmetry of left-right scattered electrons \cite{Nagaosa2010}.
Microscopically, the asymmetry is induced by the combined effect of ferromagnetic spin-polarization (Fig.~1c) and spin-orbit coupling (SOC). 
While the former breaks $\mathcal{T}$ symmetry, the later adds breaking of the 
invariance under spin rotation which, if the invariance was present, would make the AHE vanish as the invariance under $\mathcal{T}$ \cite{Suzuki2017}. 
This required symmetry breaking and associated emergent magnetic Berry curvature can arise also due to certain noncollinear antiferromagnetic structures instead of ferromagnetic moments, as predicted for Mn$_{\text{3}}$Ir \cite{Chen2014,Kubler2014}, whose magnetic lattice is shown in Fig.~1a and d. 
Large Hall conductivities were experimentally reported in related coplanar noncollinear compensated antiferromagnets Mn$_{\text{3}}$Sn \cite{Nakatsuji2015}, Mn$_{\text{3}}$Ge\cite{Nayak2016}, and Mn$_{\text{3}}$Pt\cite{Liu2018}. The nonrelativistic AHE counterpart - the topological Hall effect can occur when in the breaking of the spin-rotation invariance the SOC is replaced by a non-coplanar spin structure  as shown in certain spin-liquid candidates \cite{Machida2010}, non-coplanar antiferromagnets \cite{Shindou2001}, or skyrmions\cite{Nagaosa2013}. 

Focusing on the spin vectors and spatial configurations of magnetic atoms\cite{Suzuki2017} has naturally led to the expectation of a vanishing spontaneous Hall effect in collinear antiferromagnets\cite{Shindou2001,Surgers2014,Suzuki2017,Ghimire2018}. 
Indeed, antiferromagnets with $\mathcal{T}$ symmetry in the MLG\cite{Bradley1972,Tinkham2003}  are excluded from having spontaneous Hall effect. 
Examples encompass collinear antiferromagnets that have a symmetry termed here $\mathcal{T}_{\text{AF}}$ combining $\mathcal{T}$ and another symmetry operation, as for instance CuMnAs\cite{Smejkal2016,Godinho2018a} ($\mathcal{T}_{\text{AF}}=\mathcal{PT}$), or GdPtBi\cite{Suzuki2016} ($\mathcal{T}_{\text{AF}}=\textbf{t}_{\frac{1}{2}}\mathcal{T}$, where $\textbf{t}_{\frac{1}{2}}$ is half-unit cell translation).

The breaking of the $\mathcal{T}$ symmetry by the spin structure of ferromagnets or the more complex magnetic systems has been at the heart of all the above Hall effect considerations. In this work, we introduce an alternative relativistic spontaneous Hall mechanism.  
Here the simplified magnetic structure alone, represented by the spin vectors and spatial configurations of magnetic atoms, generates no spontaneous Hall conductivity. 
The required asymmetry is generated only when including additional atoms at noncentrosymmetric sites which can be nonmagnetic. 
Our findings demonstrate that the symmetry breaking picture based on drawing magnetic ordering as spin-projection vectors is incomplete. In general, the complete shape of magnetization density needs to be considered as we illustrate in Fig.~1e.
Here the resulting asymmetry of magnetization density and asymmetric SOC (ASOC) not only break $\mathcal{T}$ symmetry in the MLG and the spin-rotation invariance, but also generate a crystal chiral asymmetry of the left and right scattered electrons. 
We demonstrate that, apart from reversing the magnetic moments, the sign of this crystal Hall effect (CHE) can flip when reversing the crystal chirality by the rearrangement of the non-magnetic atoms while keeping the spin vectors and the positions of magnetic atoms fixed.
This is a distinct feature unavailable in common anomalous Hall systems. 
We show that the CHE mechanism allows for a large spontaneous Hall response in the room-temperature collinear antiferromagnet RuO$_{\text{2}}$ \cite{Berlijn2017a,Zhu2018}, shown in Fig.~1b. 

While RuO$_{\text{2}}$ has oxygen atoms on locally non-centrosymmetric sites, it is globally centrosymmetric. We analyse also the CHE in the quasi-two-dimensional antiferromagnet\cite{Ghimire2018}  Co$_{\frac{1}{3}}$NbS$_{\text{2}}$ which is globally non-centrosymmetric.
We catalogue all possible magnetic symmetries hosting the CHE in collinear antiferromagnets and a number of material candidates.
Finally, we discuss  the relevance of the CHE for earlier inconclusive interpretations of Hall measurements\cite{Ghimire2018,Vistoli2018} in the above mentioned Co$_{\frac{1}{3}}$NbS$_{\text{2}}$ and in the Ce-doped canted antiferromagnet CaMnO$_{\text{3}}$.

\section*{Results}
\subsection*{Crystal chirality and time-reversal symmetry breaking in collinear antiferromagnets} 
We now describe the $\mathcal{T}$ symmetry breaking due to the complex magnetization density in collinear antiferromagnets,
emphasizing the distinct nature of the CHE from the usual AHE mechanism.
The anomalous Hall conductivity in  IrMn$_\text{3}$ and similar materials is generated by the symmetry lowering due to the \textit{nontrivial} noncollinear antiferromagnetic order characterized by the local magnetic chirality\cite{Szunyongh2009,Nayak2016},
\begin{equation}
\boldsymbol\chi_{AB}^{(M)} = \textbf{S}_{A}\times \textbf{S}_{B},
\label{chiralityM}
\end{equation}
of electrons hopping along the bond connecting two noncollinear spins $\textbf{S}_{A,B}$ as shown in Fig.~1a.  
Ir Wyckoff positions are centrosymmetric and the MSG does not depend on their presence or absence in the IrMn$_{\text{3}}$ crystal. 
This justifies neglecting the nonmagnetic atoms in this class of crystals and analysing only the magnetic spin-structure \cite{Suzuki2017}, see Supplementary information (SI) Fig.~S1.
The SOC lifts the degeneracy between two magnetic states connected by spin reversals and translates the symmetry breaking into the orbital sector similarly as in the ferromagnetic AHE \cite{Szunyongh2009,Gosalbez2015,Suzuki2017}. 

From this perspective, the two-site collinear antiferromagnet, as shown in Fig.~2a, is trivial since it cannot generate any Hall signal due to the $\mathcal{T}_{\text{AF}}$ symmetry. 
However, by interlacing the magnetic lattice in Fig.~2a by the nonmagnetic atoms distributed at noncentrosymmetric positions, we can break the $\mathcal{T}_{\text{AF}}$ symmetry, as we show in Figs.~1b,e and in Fig.~2b on the rutile antiferromagnet RuO$_{\text{2}}$. 
For the collinear antiferromagnetism with quantization axis along the [100] direction, the system acquires MSG $Pn'n'm$ (Type-III), magnetic point group (MPG) $m'm'm$, and MLG $2'2'2$. The symmetry generators are $\mathcal{P}$, glide mirror plane $\mathcal{M}_{y}\textbf{t}$ ($\textbf{t}=(\frac{a}{2},\frac{a}{2},\frac{c}{2})$ marked in Fig.~2b, and antiunitary rotation $\mathcal{T}\mathcal{C}_{2z}$ and they do not change when we cant the perfectly antiparallel magnetic moments towards the [010] direction. This illustrates the ferromagnetic nature of the symmetry groups even in a fully compensated antiferromagnetic state with the Hall vector $\boldsymbol\sigma=(0,\sigma_{xz},0)$. 

In Figs.~2c,d we illustrate the microscopic mechanism which generates a non-zero Berry curvature with collinear antiferromagnetism.
In the non-magnetic state, the bands are Kramers degenerate due to the $\mathcal{P}$ and $\mathcal{T}$ symmetries\cite{Smejkal2016} of the rutile crystal. 
When we introduce the collinear antiferromagnetic order, the distribution of oxygen atoms deforms the magnetization densities around the Ru sublattices, as we show in Fig.~1e and in Fig.~S2.  
The magnetization density explicitly illustrates breaking of the $\mathcal{T}_{\text{AF}}$  symmetry for a generic crystal momentum  \textbf{k}. 
However, the effective symmetry comprising of rotating the magnetization densities (oxygen octahedra) by 90 degrees around each Ru atom in combination with half-unit cell translation enforces the two Ru atoms to be in the antiferromagnetic spin state.
In turn, the integrated even-in-magnetization quantities such as the density of states (DOS) when SOC is switched-off remain perfectly compensated (cf. Fig.~4b). 

Remarkably, the energy bands are strongly spin-split for a generic \textbf{k}, even when the relativistic SOC is switched-off in the density functional theory (DFT) calculation -- see red/blue-coloured bands in Fig.~2c. 
When the relativistic corrections are switched on, the local non-cetrosymmetricity also generates ASOC $\sim\textbf{k}\times \nabla V \cdot {\bf s}$, which additionally lowers the symmetry. 
The resulting band structure is locally spin-polarized, spin mixed, and generates the required asymmetry between left and right moving electrons as can be seen on large Berry curvature hotspots around the additional spin-splittings in Fermi surface bands shown in Fig.~2d.
The Dzyaloshinskii-Moriya interaction (DMI) generated net moment is known to be a relativistic effects of  a small magnitude \cite{IEDzyaloshinskii1958}. In contrast, our calculations demonstrate that the spin-symmetry breaking is not a correction but a strong effect reflected in a large magnitudes of the intrinsic Hall conductivity.

We calculate the intrinsic  Hall conductivity (independent of disorder-scattering) by integrating the Berry curvature,
$
\bold{\Omega}(\textbf{k})=-\text{Im} \langle\partial_{\textbf{k}}u(\textbf{k}) \vert \times \vert \partial_{\textbf{k}}u(\textbf{k}) \rangle,
$
in the crystal momentum space (see Methods). 
In Fig.~S3 we show that the non-vanishing integral component $\int dk_{x}\Omega_{y}(\textbf{k})$ is even in $k_{y}$ as we expect from the symmetry analysis, while  the $\mathcal{M}_{y}$, $\mathcal{P}$ and $\mathcal{TC}_{2z}$ symmetries imply that $\int dk_{x}\Omega_{x}(\textbf{k})=0$, and $\mathcal{M}_{y}$,  and $\mathcal{TC}_{2z}\mathcal{M}_{y}$ yield $\int dk_{z}\Omega_{z}(\textbf{k})=0$. We obtain  $\sigma_{xz}$ = 35.7~Scm$^{-1}$, demonstrating a large crystal Hall conductivity in RuO$_{\text{2}}$ at the charge neutrality point. The DFT calculations of the CHE are extensively discussed below.

\subsection*{Crystal chirality control of the Hall conductivity}
We can illustrate the crystal chirality features on a simplified model of a collinear antiferromagnet with CHE.
Inspired by the Haldane's quantum anomalous Hall effect model \cite{Haldane1988}, we have found a minimal Hamiltonian simultaneously hosting the staggered antiferromagnetic potential in combination with $\mathcal{T}$ symmetry breaking compatible with the existence of the Hall conductivity:
\begin{equation}
H=t\sum_{ij}c_{i}^{\dagger}c_{j}+J_{n}\sum_{i}\textbf{u}_{i}\cdot\textbf{s} c_{i}^{\dagger}c_{i} + \lambda\sum_{ij} \hat{\boldsymbol\chi}_{ij}^{(C)}\cdot \textbf{s} c_{i}^{\dagger}c_{j}\,.
\label{soc}
\end{equation} 
Here the first term describes electron hopping between nearest neighbour magnetic sites $i$ and $j$ on a body center cubic lattice and the second term describes on-site exchange field with an alternating direction on the neighbouring sites, $\textbf{u}_{i}=-\textbf{u}_{j}$ ($\textbf{s}$ are spin Pauli matrices).  
These two terms represent a  tight-binding model of a collinear antiferromagnet with the MPG $\mathcal{T}$ symmetry. 
We lower the symmetry of the Hamiltonian (\eqref{soc}) by a staggered ASOC term 
due to the local crystal chirality defined as
\begin{equation}
\boldsymbol\chi^{(C)}_{AB}= \textbf{d}_{A}\times\textbf{d}_{B} ,
\label{chiralityC} 
\end{equation}
arising due to the non-magnetic atoms at the non-centrosymmetric positions. Here $\textbf{d}_{AO}$ and $\textbf{d}_{OB}$ are vectors connecting two nearest-neighbour Ru atoms with the common interlaced oxygen atom (cf. Fig.~2b), where $\hat{\boldsymbol\chi}_{ij}^{(C)}$ in Eq.~\eqref{hamiltonian} marks the chirality unit vector.  

While the exchange and ASOC  in Eq.~\eqref{soc} separately do not break the $\mathcal{T}$ symmetry, their combination does break the MPG $\mathcal{T}$-symmetry. 
The model crystal momentum Hamiltonian can exhibit nodal-chain band topology\cite{Bzdusek2016} and a large Berry curvature along the [010] direction and we discuss the details in Fig.~S4. In contrast to the Haldane's quantum anomalous Hall model,
our model demonstrates the possibility of the spontaneous Hall conductivity without the necessity for ferromagnetism or complicated non-collinear and non-coplanar antiferromagnetism, even in a globally centrosymmetric system.

We now demonstrate the possibility to control the Hall conductivity sign by swapping the crystal chirality. In Figs.~3a,b we show the RuO$_{\text{2}}$ crystal with the two possible distributions of the oxygen atoms corresponding to the opposite crystal chiralities $\boldsymbol\chi^{(C)}=\pm 1$. While the MSG is the same in both cases, the local magnetization densities, obtained from the DFT calculations, are rotated by 90 degrees\cite{Gopalan2011}. 
In Fig.~3c we plot the energy bands corresponding to the crystal in Fig.~3a. The red and blue arrows mark spin up and down projection for the bands calculated without SOC. When we include the SOC we obtain additional splittings of the bands and a large Berry curvature, as we show in Figs.~2d and 3d. The red and blue colours correspond to the opposite local chirality crystals shown in Figs.~3a,b.  

The flipping of the sign of CHE $\sigma_{xz}$  with the N\'{e}el vector reversal is consistent with the Onsager relations. 
The two crystals in Figs.~3a,b can be mapped on each other by the $\mathcal{T}$ operation combined with a half-unit cell translation and this symmetry ensures the same magnitude, while opposite sign, of $\sigma_{xz}$ for the two crystal chiralities. We can see this also from our minimal model analysis,  where $\boldsymbol\sigma$ changes sign when the chirality $\boldsymbol\chi_{ij}^{(C)}$ in the spin-orbit term is reversed. 
From this, we can draw a comparison to the AHE in ferromagnets and non-collinear antiferromagnets, where the sign reversal of the Hall conductivity is governed by the reversal of the net magnetic moment \textbf{m} or the magnetic chirality \eqref{chiralityM}. The sign of the CHE in the simple collinear antiferromagnets such as RuO$_{\text{2}}$ is, instead, determined by the sign of the dot-product $\textbf{n}\cdot\boldsymbol\chi_{ij}^{(C)}$ underlining the crystal mechanism of the symmetry breaking in this collinear antiferromagnet.

\subsection*{Crystal Hall phenomenology in RuO$_2$}
In Fig.~4a we identify a sizable CHE conductivity in the room temperature collinear antiferromagnet RuO$_{\text{2}}$ by our first-principle calculations. For artificially constrained perfectly antiparallel spin moments along the [100] axis, $\boldsymbol{\sigma}\parallel [010]$ and we obtain $\sigma_{xz}$ = 36.4~Scm$^{-1}$. For a canting angle $\approx1^{\circ}$ obtained from the DFT calculation, $\textbf{m}\parallel [010]$ and $\sigma_{xz}$ = 35.7~Scm$^{-1}$. 
Note that among the rutile antiferromagnets \cite{Tinkham2003,Bradley1972}, a metallic phase is rare which makes the recently discovered  \cite{Berlijn2017a,Zhu2018} itinerant antiferromagnetism in 
RuO$_{\text{2}}$ exceptional within this family of simple collinear antiferromagnets. Our DFT calculations (see Fig.~S5) show that for a medium strength Hubbard parameter ($U\sim 1- 3$ eV), antiferromagnetism and metallic density of states (DOS) coexist, consistent with previous reports \cite{Berlijn2017a,Zhu2018}. 
We set in all plots in the main text U$\sim$ 2~eV, which reproduces best the experimental antiferromagnetic moments.

When turning the sizable SOC off in our DFT calculation, we observe a perfect antiferromagnetic compensation in the Ru-projected DOS. With the large atomic SOC turned on, only minute corrections to the DOS occur, as shown in Fig.~4b. They result in a small net magnetic moment, $\textbf{m}=\textbf{m}_{A}+\textbf{m}_{B}$, of a magnitude $\sim0.05$~$\mu_{\rm B}$ due to the DMI \cite{IEDzyaloshinskii1958}. 
Here $\textbf{m}_{A/B}$ are magnetizations of the antiferromagnetic $A$ and $B$ sublattices. In comparison, the N\'eel vector $\textbf{n}=(\textbf{m}_{A}-\textbf{m}_{B})/2$ has a magnitude $\sim1.17$~$\mu_{\rm B}$. 

To gain further insight, we calculate the dependence of the CHE for $\textbf{n}\parallel[100]$ on the canting angle between magnetizations of sublattices $A$ and  $B$, see Fig.~4a. 
Furthermore, we separate  in Fig.~4a $\sigma_{xz}$ into a contribution even in $\textbf{m}$:
\begin{equation}
\sigma_{xz}^{\text{CHE}}=[\sigma_{xz}(\textbf{n},\textbf{m})+\sigma_{xz}(\textbf{n},-\textbf{m})]/2,
\end{equation}
and odd in  $\textbf{m}$: 
\begin{equation}
\sigma_{xz}^{\text{AHE}}=[\sigma_{xz}(\textbf{n},\textbf{m})-\sigma_{xz}(\textbf{n},-\textbf{m})]/2.
\end{equation}
Here $\sigma_{xz}^{\text{AHE}}$ corresponds to a contribution induced by the small net moment, analogous to the AHE in ferromagnets. 
We see that this term is roughly linear in $\textbf{m}$ (shown in Fig.~4c), at least for $\vert\phi\vert\lesssim10^{\circ}$, while $\sigma_{xz}^{\text{CHE}}$ is almost constant at small $\phi$ and dominates the contribution to $\sigma_{xz}$.
Hence the small net magnetic moment has a negligible effect on $\sigma_{xz}$. 
This is in striking contrast to the recently studied antiferromagnets GdPtBi\cite{Suzuki2016} and EuTiO$_{\text{3}}$\cite{Takahashi2018}, which order in a $\mathcal{T}$-invariant MPG and whose observed AHE is entirely due to the canting induced by an applied external magnetic field.

In Fig.~4d we plot the intrinsic crystal Hall conductivity for the N\'eel vector orientation along [100] and [110] crystal axies
as a function of the Fermi level position which simulates, e.g., alloying.  We observe peak values of $\sim1000$~S/cm, corresponding to record magnitudes reported for the AHE in ferromagnets or non-collinear antiferromagnets  \cite{Nagaosa2010,Nayak2016} (see Tab.~S1). 
This illustrates also a large anisotropy of the crystal Hall conductity\cite{Roman2009} which can be understood in terms of  the hybridization of linear band crossings and the gapping of nodal-line features\cite{Kim2018d} in dependence on the N\'{e}el vector orientation \cite{Smejkal2016} (see Fig.~6). The MSG $Pnn^\prime m^\prime$ for $\textbf{n}\parallel [100]$ changes to the MSG $Cnn^\prime m^\prime$ for $\textbf{n}\parallel [110]$.

In Fig.~S7c,d, we observe that DMI generates a small magnetization that is perpendicular to the N\'eel vector when $\textbf{n}\parallel[100]$ while, for $\textbf{n}\parallel[110]$, it generates a small parallel magnetization. While in the former case the Hall vector is perpendicular to the N\'eel vector, in the latter case the two vectors are parallel as we schematically illustrate in Fig.~4e. 
Also, from Figs.~4a,c, and Fig.~S7 we see that the crystal Hall conductivity is proportional to neither spin nor orbital magnetization, and, for a generic angle of the N\'eel vector, the mutual orientation of the N\'eel and Hall vectors is arbitrary and depends on microscopic details.

\subsection*{Proposals for the experimental observation of the crystal Hall effect}
To measure the CHE in RuO$_{\text{2}}$ we need to ensure one dominating crystal chirality (e.g. by growing single-domain samples) and the correct orientation of the N\'eel vector. Preferably we suggest to orient the crystal growth direction along the Hall vector, two possibilities are marked in Fig.~4e. 
We note that the easy axis in RuO$_{\text{2}}$ can point along the [001] direction (as experimentally observed in bulk\cite{Berlijn2017a} and consistent with our calculations for stoichiometric RuO$_{\text{2}}$ shown in Fig.~S5) or slightly tilted from the [001] axis as reported for thin films \cite{Zhu2018}. Any nonzero tilting of the easy axis from the [001] direction is sufficient to generate nonzero CHE.

Otherwise\cite{Berlijn2017a}, an additional spin-flop field can be applied along the [001]-axis to force the N\'eel vector towards the (001)-plane. 
Alternatively, our DFT calculations show (see Fig.~S5) that by a few per cent alloying, e.g. in $\text{Ru}_{\text{1+x}}\text{O}_{\text{2-x}}$ or $\text{Ru}_{\text{1-x}}\text{Ir}_{\text{x}}\text{O}_{\text{2}}$, the easy axis can be constrained to  the (001) plane. 
With the N\'eel vector in the (001)-plane, an in-plane magnetic field can be applied to select one of the two domains with opposite in-plane N\'eel vectors and corresponding opposite signs of the Hall effect, consistent with domain energies in Fig.~4f. 
 
This also demonstrates the possibility to turn the CHE on and off by reorienting the N\'eel vector. 
In contrast, the AHE in ferromagnets is allowed by symmetry for any direction of the magnetization. 

The sign of the Hall conductivity can be controlled also by the global crystal chirality. We explain this on the Co$_{\frac{1}{3}}$NbS$_{\text{2}}$ crystal (its low-symmetry magnetization isosurfaces are shown in Fig.~5a), the quasi-2d hexagonal collinear antiferromagnet derived from the Van der Waals crystal of transition metal dichalcogenide NbS$_{\text{2}}$ \cite{Ghimire2018}. 
The opposite sign of crystal Hall conductivity, shown in Figs.~5b,c, corresponds to the two crystals with the opposite sense of the spatial inversion symmetry breaking, marked $L$ and $R$ in Fig.~5b.

Co$_{\frac{1}{3}}$NbS$_{\text{2}}$, with collinear antiferromagnetic moments\cite{Parkin1983}, has the $C2^\prime 2^\prime 2_{1}$ MSG and the same MLG as RuO$_{\text{2}}$  ($2'2'2$), where the unprimed rotational axis $\mathcal{C}_{2}$ is perpendicular to the hexagonal layers and $\boldsymbol\sigma\parallel \textbf{a}_{\mathcal{C}_{2}}$ (according to our classification in Tab. 1). 
However, the global $\mathcal{P}$ symmetry breaking promotes the role of ASOC, as we show in Fig.~5d, where the bands are split along the high symmetry axes, not only at high symmetry points (see details of the energy bands in Fig.~S8). 
The energy bands, e.g. around the $H$ point, are substantially split by the ASOC and in combination with collinear antiferromagnetism, a large Berry curvature $\Omega_{z}$ is generated as we illustrate in the Fig.~S8 on the Berry curvature summed up to the lowest energy band shown in Fig.~5d. The Berry curvature appears to be concentrated around these antiferromagnetic generalisations of Kramers-Weyl-like dispersions\cite{Chang2018}.

We note that the  spontaneous Hall effect recently detected in Co$_{\frac{1}{3}}$NbS$_{\text{2}}$ \cite{Ghimire2018} could not be reconciled with a collinear antiferromagnetic order inferred from neutron scattering\cite{Parkin1983}. Our first-principles calculations shown in Fig.~5c give a magnitude of the CHE in hole-doped (Fermi energy $\sim -0.75 eV$) Co$_{\frac{1}{3}}$NbS$_{\text{2}}$ which is consistent with the experimental value\cite{Ghimire2018} (27~S/cm).

\section*{Discussion}
While the symmetry allowed direction of the Hall vector $\boldsymbol\sigma$ depends only on the MLG, the possibility to control the sign of the CHE  by the local or global crystal chirality depends on the full MPG.  
To enumerate all possible symmetries allowing for the CHE in collinear antiferromagnets we start by excluding antiferromagnetic symmetries incompatible with the existence of a Hall vector. Among those are all MSG-type-IV antiferromagnets with $\mathcal{T}_{\text{AF}}=\textbf{t}_{\frac{1}{2}}\mathcal{T}$ symmetry ($\textbf{t}_{\frac{1}{2}}$ is half-unit cell translation as e.g. in GdPtBi), and MSG-type-III antiferromagnets $\mathcal{T}_{\text{AF}}=\mathcal{PT}$ symmetry  (e.g. CuMnAs, or Mn$_{\text{2}}$Au) which have the $\mathcal{T}$ symmetry in the MLG. In total 275 MSG/31 MPGs/10 MLGs of type I and III remain as candidates for spontaneous Hall effects.
However, simple collinear antiferromagnetism is not compatible with 3-fold, 4-fold, and 6-fold rotational symmetries \cite{Turov1965}. We summarize in Tab.~1 the remaining 12 MPGs (and 4 MLGs) that may host the CHE in collinear antiferromagnets. 

We can formulate simple rules allowing for a fast determination of the orientation of the Hall vector $\boldsymbol\sigma$ based on the existence of these only 4 MLGs. (1.) In MLG $1$ the orientation of $\boldsymbol\sigma$ is arbitrary and depends on microscopic details of the electronic structure. (2.) In the systems with $2'$ rotational axis the Hall vector is perpendicular to the axis and the orientation within this plane is set microscopically. (3.) The $2$ fold rotational axis constrains the Hall vector to be parallel to this axis (see Fig.~1a, and 5b) and the orientation of the Hall vector is determined uniquely by the symmetry. All the remaining possibilities can be derived from these 3 (for instance in $2'2'2$ the Hall vector is perpendicular to both $2'$ and parallel to $2$).  
 
We point out that as many as $\sim 10\%$ of the total of $\sim 700$ magnetic structures reported in the Bilbao MagnData database\cite{Gallego2016} belong to the class of collinear antiferromagnets in which the CHE is allowed by symmetry.  
We point out that our CHE mechanism will materialize in these candidates possibly also in its optical or thermal variants\cite{Smejkal2018a}.
In Tab.~1 we list some additional material candidate examples such as orthoferrites, perovskites, or corundum structure materials. 
In addition, we provide in the SI a classification table Tab.~S2 of all 31 MPG allowing for Hall effects also in non-collinear spin-structures. 

The CHE might also contribute to Hall signals which were earlier taken as a signature of nontrivial and topological magnetization textures.
This applies, e.g., to the measured spontaneous Hall signal in a Ce-doped canted antiferromagnet CaMnO$_{\text{3}}$ (MPG $2^\prime/m^\prime$) \cite{Vistoli2018}. Apart from the AHE contribution due to the net magnetic moment, our symmetry analysis shows that the CHE associated with the N\'{e}el vector, rather than the canting moment (cf. Figs.~4a,c) is allowed in this material due to the oxygen non-centrosymmetric positions. The spikes arising in the Hall signal by applying a magnetic field can be alternatively explained as a convolution of two spontaneous Hall signals from material regions with the opposite Hall sign \cite{Gerber2018}. These two regions might correspond to the two crystallites with the opposite sign of the CHE. 
Furthermore, methods for growing single-crystal-chirality systems\cite{BanerjeeGhosh2018} can be used to enhance the Hall signal.

Finally, we remark that existing mechanisms of the quantum spontaneous Hall effect rely either on rare ferromagnetic insulators or on fragile diluted magnetic topological insulators with low critical temperatures and small magnetic band-gaps \cite{Tokura2019}. Our crystal spontaneous symmetry breaking represents a long-sought mechanism marrying strong Hall response with a robust room-temperature intrinsic collinear antiferromagnetism\cite{Smejkal2018a}.

\section*{Methods}
The Hamiltonian of our model \eqref{soc} in the crystal-momentum space reads,
\begin{eqnarray}
H_{\textbf{k}}&=-4t\tau_{x}\cos\frac{k_{x}}{2}\cos\frac{k_{y}}{2}\cos\frac{k_{z}}{2}+ \tau_{z}J_n\textbf{s} \cdot\textbf{n}+\nonumber \\ 
&4i\lambda\tau_{x}\sin\frac{k_{z}}{2}\left[ s_{xy}^{(-)}\sin\frac{k_{x}+k_{y}}{2} +s_{xy}^{(+)}\sin\frac{k_{x}-k_{y}}{2}  \right] \,,
 \label{hamiltonian}
\end{eqnarray}
where $\boldsymbol\tau$ are site Pauli matrices and  $s_{xy}^{(\pm)}=s_{x}\pm s_{y}$.  

The MSG of rutile antiferromagnets for the N\'{e}el vector along the [100] and [110] axis, respectively, are:
\begin{eqnarray*}
Pnn^\prime m^\prime &:& \mathcal{P}, \mathcal{G}_{y}, \mathcal{S}_{2y},\mathcal{TC}_{2z},\mathcal{TM}_{z},\mathcal{TG}_{x},\mathcal{TS}_{2x}, \\
 Cmm^\prime m^\prime &:& \mathcal{P},\mathcal{M}_{xy},\mathcal{C}_{2xy}, \mathcal{TC}_{2z},\mathcal{TM}_{z},\mathcal{TM}_{\overline{x}y},\mathcal{TC}_{2\overline{x}y} \nonumber,  
 \label{sym}
\end{eqnarray*}
where the non-symmorphic symmetries are the unitary glide plane $\mathcal{G}_{y}=\mathcal{M}_{y}\textbf{t}_{\frac{1}{2}}$, screw rotation $\mathcal{S}_{2y}=\mathcal{C}_{2y}\textbf{t}_{\frac{1}{2}}$, and anti-unitary $\mathcal{TG}_{x}=\mathcal{TM}_{x}\textbf{t}_{\frac{1}{2}}$, and $\mathcal{TS}_{2x}=\mathcal{TC}_{2x}\textbf{t}_{\frac{1}{2}}$ and in the case of $Cmm^\prime m^\prime$ there are also operations coupled by $\textbf{t}=(1/2,1/2,0)$. 

The MSG $P4_2'/mnm'$ (and the corresponding MPG  $4'/mm'm$) of rutile antiferromagnets for the N\'{e}el vector along the [001] prohibits the existence of Hall vector. 

The MSG of Co$_{\frac{1}{3}}$NbS$_{\text{2}}$ for the N\'{e}el vector along the [100] is $C2^\prime 2^\prime 2_{1}$ and includes symmetry operations $\mathcal{TC}_{2x},\mathcal{TS}_{2y}$, and $\mathcal{S}_{2z}$, where $\textbf{t}=(0,0,1/2)$ and all the symmetries coupled by $\textbf{t}=(1/2,1/2,0)$. 
The MLG (MPG stripped off inversions) is the same, $2^\prime 2^\prime 2$, for both RuO$_{\text{2}}$, and Co$_{\frac{1}{3}}$NbS$_{\text{2}}$ crystal, and thus also the shape of the conductivity tensor.

The time reversal operation acts on the Berry curvature as $\mathcal{T}\bold{\Omega}(\textbf{k})=-\bold{\Omega}(-\textbf{k})$, and the following symmetries operate as: 
\begin{eqnarray*}
\mathcal{P}\bold{\Omega}(\textbf{k})&=&\bold{\Omega}(-\textbf{k}),  \\
\mathcal{M}_{y}\bold{\Omega}(\textbf{k})&=&(-\Omega_{x},\Omega_{y},-\Omega_{z})(k_{x},-k_{y},k_{z}), \\
\mathcal{C}_{2y}\bold{\Omega}(\textbf{k})&=&(-\Omega_{x},\Omega_{y},-\Omega_{z})(-k_{x},k_{y},-k_{z}),   \\
\mathcal{TC}_{2z}\bold{\Omega}(\textbf{k})&=&(\Omega_{x},\Omega_{y},-\Omega_{z})(k_{x},k_{y},-k_{z}),  \\
\mathcal{TM}_{z}\bold{\Omega}(\textbf{k})&=&(\Omega_{x},\Omega_{y},-\Omega_{z})(-k_{x},-k_{y},k_{z}),  \\
\mathcal{TM}_{x}\bold{\Omega}(\textbf{k})&=&(-\Omega_{x},\Omega_{y},\Omega_{z})(k_{x},-k_{y},-k_{z}),  \\
\mathcal{TC}_{2x}\bold{\Omega}(\textbf{k})&=&(-\Omega_{x},\Omega_{y},\Omega_{z})(-k_{x},k_{y},k_{z}) .
\label{syms} 
\end{eqnarray*}

We calculated the Hall conductivity in our model and in DFT as,
\begin{equation}
\sigma_{xz}=-\frac{e^{2}}{\hbar}\int\frac{d\textbf{k}}{(2\pi)^{3}}\sum_{n}f(\textbf{k})\Omega_{y}(n,\textbf{k}),
\label{Eq_Berry}
\end{equation}
where the Berry curvature, $\Omega_{y}(n,\textbf{k})$, is defined as in the main text for each individual band with the quantum number $n$ and corresponding Bloch functions $u_n(\textbf{k})$, and $f(\textbf{k})$ is the Fermi-Dirac distribution\cite{Vanderbilt2016b}. 

We have calculated the electronic structure of RuO$_{\text{2}}$ in the full potential DFT code ELK\cite{elk} and pseudopotential DFT code VASP \cite{Kresse1993,*Kresse1994,*Kresse1996,*Kresse1996a}, within PBE+U+SOC (spherically invariant version of DFT+U) \cite{Kresse1994a,*Joubert1999} on a $12\times 12\times 16$ $k$-point grid and we use energy cut-off 500~eV. In the case of RuO$_{\text{2}}$ crystal, we performed our magnetocrystalline anisotropy calculations on a $16\times 16\times 24$ $k$-point grid. For the transport calculations, we used only VASP results. We obtained Wannier functions on a $12\times 12\times 16$ grid using the Wannier90 code\cite{Mostofi2014} and we calculated the Hall conductivity by employing Eq.~(\ref{Eq_Berry}) evaluated on the Monkhorst-Pack grid with a $321^{3}$ crystal momentum integration mesh in Wannier tools \cite{WU2017}. In the case of Co$_{\frac{1}{3}}$NbS$_{\text{2}}$, we evaluate Wannier functions on a $12\times 12\times 6$ grid, and we use $241^{2}\times 201$ crystal momentum points for the Hall conductivity calculation. 
We tested our first principles calculation methodology on the experimentally and theoretically investigated anomalous Hall conductivity in ferromagnets and non-collinear antiferromagnets and we obtained an agreement with the previous reports (e.g., 234 S/cm in IrMn$_{\text{3}}$, cf. Ref.~\onlinecite{Chen2014}). 

The distortion of the tetragonal unit cell (used in calculations with N\'{e}el vector along the [100] axis) with lattice parameters, $a=4.528$, $b=4.536$, $c=3.124$\r A, due to the magneto-elastic coupling does not change magnetic symmetries and we use in the main text tetragonal unit cell. For $\protect \textbf {n} \parallel [110]$, and [001] we obtained from our DFT calculations after relaxation : $a=b=4.5337$, $c=3.124$\r A, and $a=b=4.5331$, $c=3.1241$\r A, respectively, consistent with a previous report\citep{Berlijn2017a}. 

\begin{figure*}[h]
\centering
\includegraphics[width=1\textwidth]{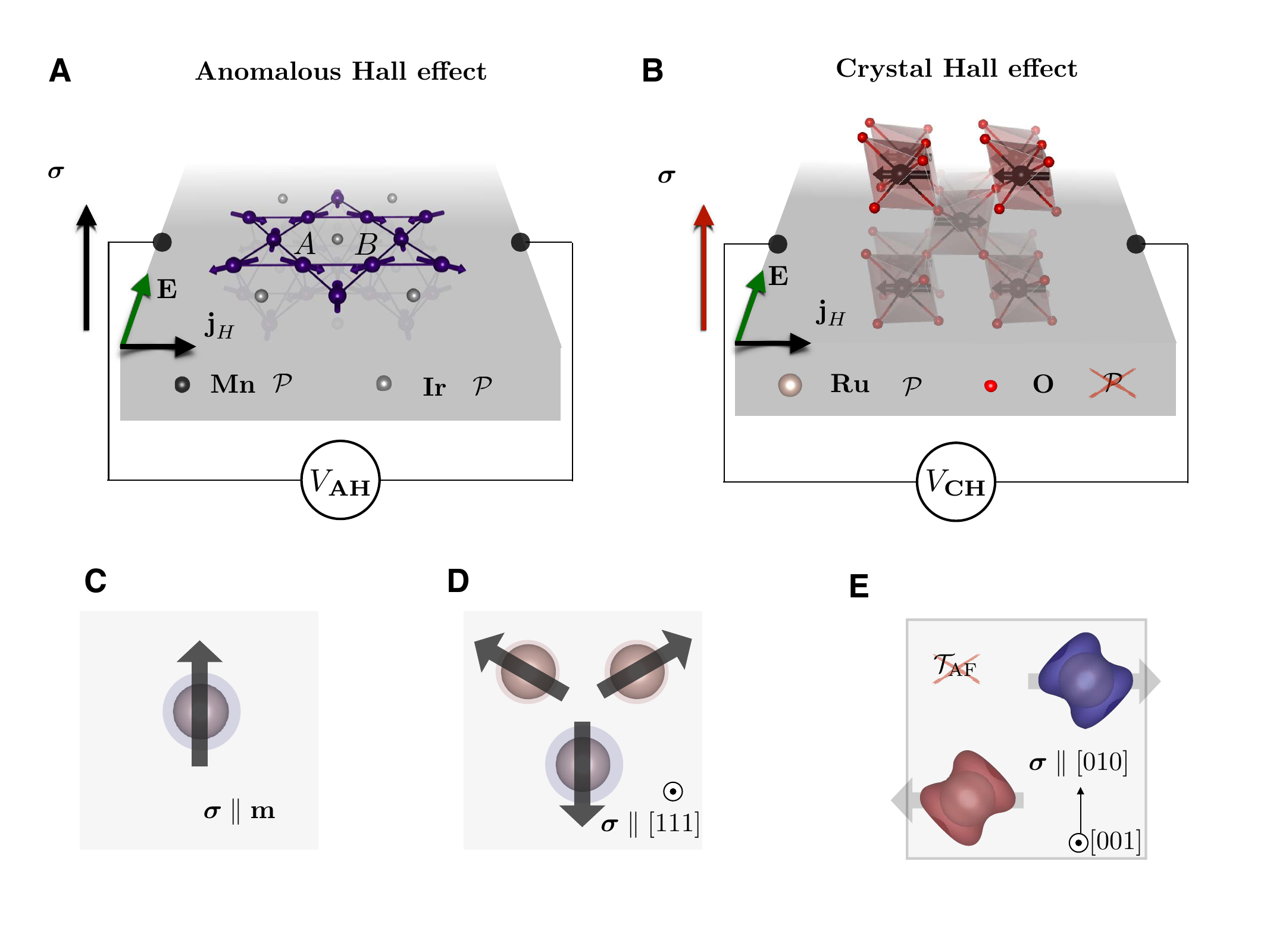}
\caption{{\bf Anomalous and crystal Hall effects and magnetization isosurfaces.}
{\bf (A)} Anomalous Hall effect due to a Hall vector ($\boldsymbol\sigma$) generated by the magnetic chirality of the non-collinear antiferromagnetic order (purple arrows) in Mn$_{\text{3}}$Pt (Mn: dark spheres, Pt: grey spheres). Mn and Pt atoms occupy centrosymmetric sites. Magnetic moments on sites $A$ and $B$ have a non-zero local magnetic chirality $\boldsymbol\chi^{(M)}_{AB}$.
{\bf (B)} Crystal Hall effect due to the Hall vector generated by the local crystal chirality $\boldsymbol\chi^{(C)}_{AB}$  (explained in Fig.~2)
(Ru: light brown spheres, O: red spheres) with a collinear antiferromagnetic order (black arrows). While the crystal has inversion centre at the Ru atom, the O atoms are at noncentrosymmetric positions.
The conventional symmetry breaking mechanism in anomalous Hall effect in ferromagnets {\bf (C)} ($\textbf{m}$ marks magnetization vector) or noncollinear antiferromagnets {\bf (D)} can be captured by the spin structure of the magnetic ions only (black arrows). 
We plot in panels {\bf (C - E)} also schematic of first-principle calculated magnetisation density isosurfaces with projection along [100] direction. In the case of crystal Hall antiferromagnet {\bf (E)}  complete magnetization density shape is required to capture the spontaneous symmetry breaking.}
\end{figure*}

\begin{figure*}[h]
\centering
\includegraphics[width=1\textwidth]{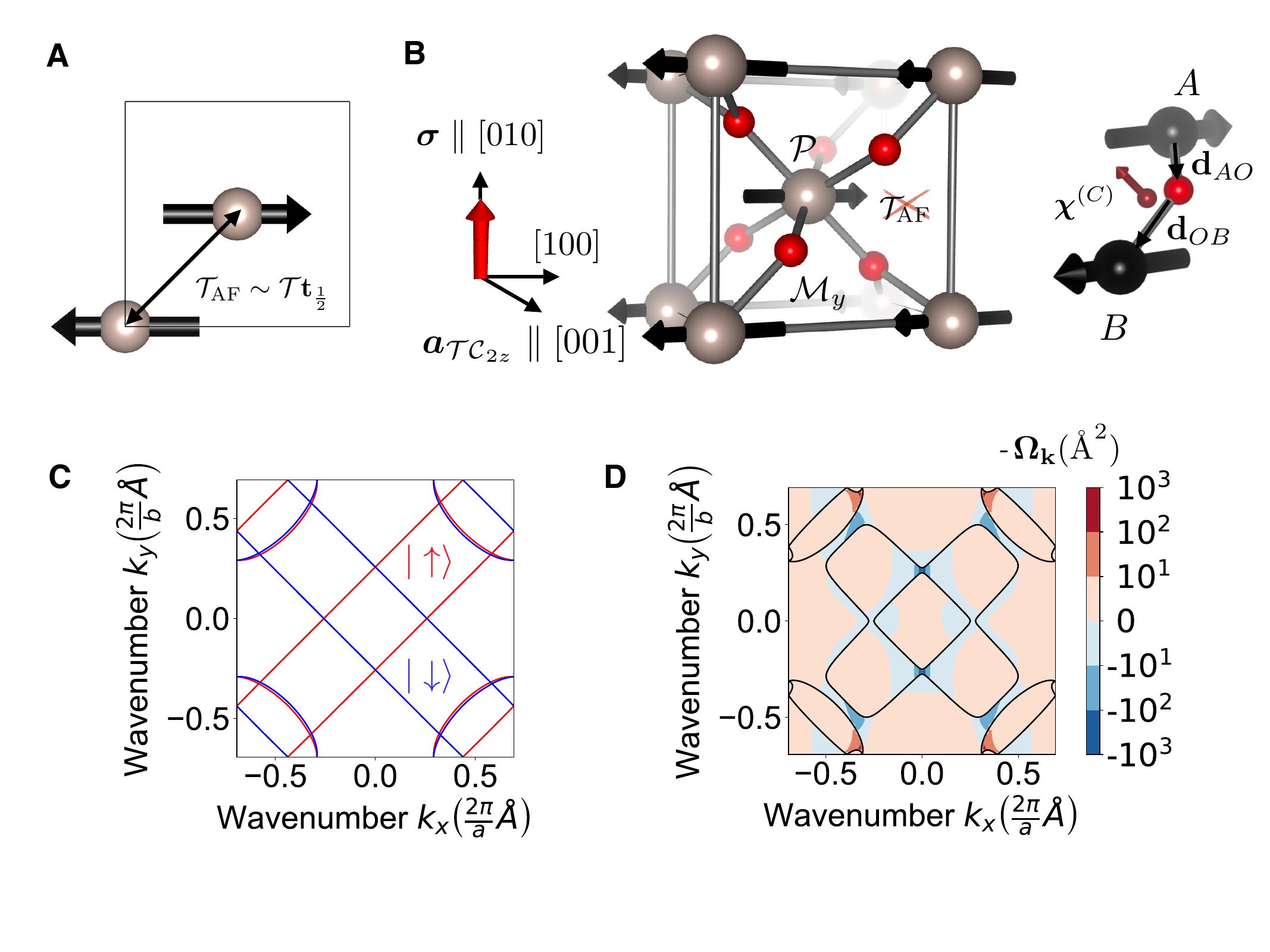}
\caption{{\bf Spontanous crystal chiral symmetry breaking and spin assymetric Fermi surfaces.}
{\bf (A)} Collinear antiferromagnet with effective time-reversal symmetry $\mathcal{T}\textbf{t}_{\frac{1}{2}}$.
{\bf (B)} Left: the unit cell of antiferromagnetic RuO$_{\text{2}}$ with the N\'eel vector along the [100]-axis and and marked crystal symmetries. Right: detail of generation of local crystal chirality by noncentrosymmetric oxygen atoms $\boldsymbol\chi^{(C)}_{AB}\sim\textbf{d}_{AO} \times \textbf{d}_{OB}$.
{\bf (C)} Antiferromagnetic Fermi surface cut at wavevector $k_{z}=0$ calculated without spin-orbit coupling. The spin up and down projections are coloured in red and blue. 
{\bf (D)} Calculations with spin-orbit coupling of crystal momentum resolved intrinsic Hall conductivity and Berry curvature $-\Omega_{y}(k_{x},k_{y},0)$ in atomic units.}
\end{figure*}

\begin{figure*}[h]
\centering
\includegraphics[width=1\textwidth]{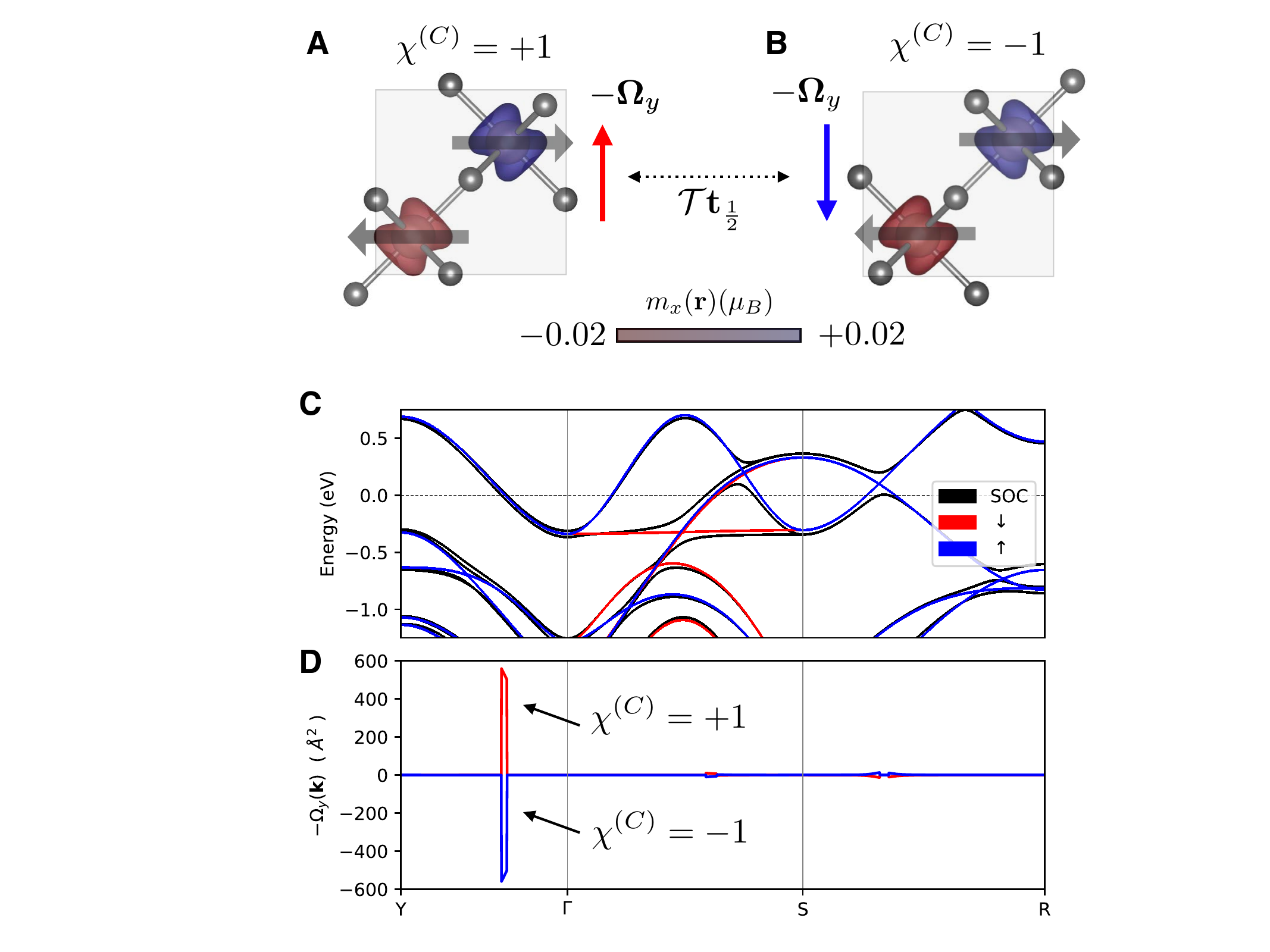}
\caption{{\bf Crystal chirality control of Hall conductivity sign.}
View along the tetragonal crystal axis on the RuO$_{\text{2}}$ crystal with two possible configurations of nonmagnetic oxygen atoms {\bf (A)} and {\bf (B)}. Redistribution of the oxygen atoms does not change the magnetic symmetry of the crystal, however, changes the local crystal chirality orientation $\boldsymbol\chi^{(C)}$ and rotates by 90 degrees the shape of the magnetization density isosurfaces. 
 {\bf (C)} Calculated energy bands in RuO$_{\text{2}}$ antiferromagnets without (red and blue corresponds to the spin-projections), and with (black) spin-orbit coupling. 
  {\bf (D)} Largest contribution to the Berry curvature  $ -\boldsymbol\Omega$ originates from the spin-split bands by the spin-orbit coupling. 
  The red and blue colour corresponds to the two opposite crystal chiralities $\boldsymbol\chi^{(C)}$ and demonstrates the expected Berry curvature sign change (compare to panel {\bf(A)}). 
}
\end{figure*}

\begin{figure*}[h]
\centering
\includegraphics[width=1\textwidth]{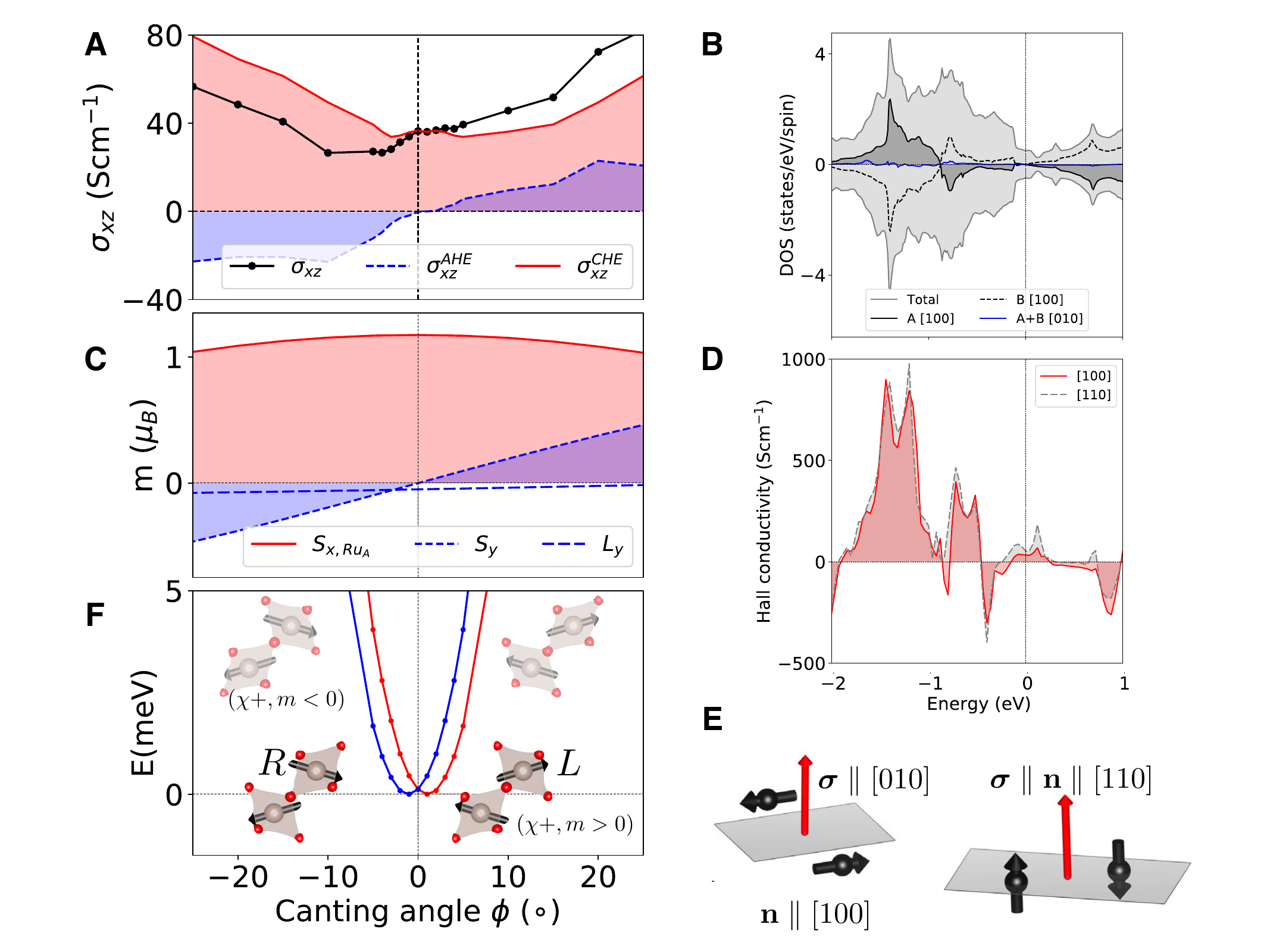}
\caption{{\bf First-principle calculation of sizable and  anisotropic Crystal Hall effect in RuO$_{\text{2}}$.}
{\bf (A)} The dependence on the canting angle of the Hall conductivity and its separation into the anomalous (ferromagnetic) and crystal (antiferromagnetic) parts. 
 {\bf (B)}  Ru sublattice $A$ (solid) and $B$  (dashed) projected DOSs for the N\'eel vector along the [100] axis. Black solid and dotted lines show calculations with spin-orbit coupling of the DOS component for moments projected along the [100] axis. Blue line shows the sum of sublattice DOSs for the moment projection along the [010] axis which  corresponds to the small canting of the antiparallel moments due to Dzyaloshinskii-Moriya interaction.
{\bf (C)} The dependence on the canting angle of the spin $S_{x}, S_{y}$ (along x projected on single Ru atom, along y total net spin moment) and orbital magnetization $L_{y}$.
 {\bf (D)} Energy dependence of the calculated crystal Hall conductivity for $\textbf{n}\parallel [100]$ (red solid line) and $\textbf{n}\parallel [110]$ (gray dashed line). 
{\bf (E)} The mutual orientation of the  N\'eel vector $\textbf{n}$, and Hall vector $\boldsymbol\sigma$. 
{\bf (F)} Two magnetic domains with opposite N\'eel vector induced by opposite field $H$ and the corresponding energy costs for canting. $\textbf{H}\parallel$ [010] corresponds to canting angles $\phi >0 $ and prefers  $\textbf{n}  \parallel [100] $ (red) over  $\textbf{n} \parallel [\overline{1}00]$ (blue). In the inset are depicted four combination of local crystal chirality and N\'{e}el vector orientations. The two marked $L$ and $R$ have the lowest energy.
}
\end{figure*}

\begin{figure*}[h]
\centering
\includegraphics[width=1\textwidth]{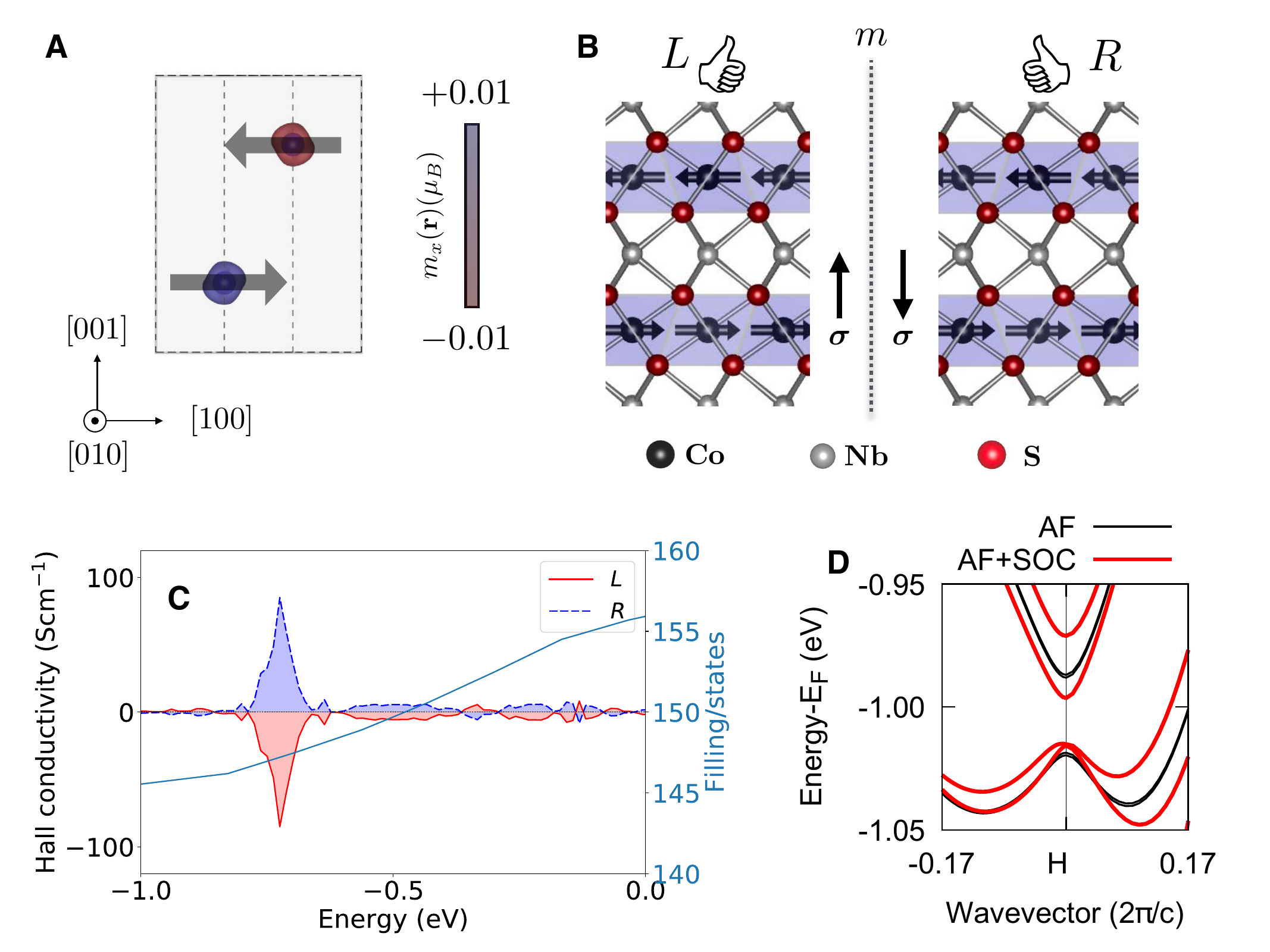}
\caption{{\bf Crystal Hall conductivity due to the global crystal chirality in Co$_{\frac{1}{3}}$NbS$_{\text{2}}$ antiferromagnet.}
{\bf (A)} The calculated magnetization isosurfaces in Co$_{\frac{1}{3}}$NbS$_{\text{2}}$ antiferromagnet exhibit low symmetry and illustrate the global chiral symmetry breaking.
{\bf (B)} The crystal of Co$_{\frac{1}{3}}$NbS$_{\text{2}}$ antiferromagnet ($"L"$, with a left-handed chirality) and its mirror $m$ image ($"R"$, with a right-handed chirality). Note that the mirror $m$ maps the two chiralities onto each other by redistributing the nonmagnetic S atoms, while preserving the magnetic atoms positions and colllinear antiferromagnetism orientation.
{\bf (C)} The calculated crystal Hall conductivity (left axis) changes sign when the crystal chirality is reversed from left- to right-handed. The right axis corresponds to the calculated electron filling in dependence on energy.
{\bf (D)} Bandstructure detail of antiferromagnetic Co$_{\frac{1}{3}}$NbS$_{\text{2}}$ without (black line) and with (red) spin-orbit coupling. We show fraction of the $LHA$ path in Brillouin zone.}
\end{figure*}

\begin{table*}[ht]
\begin{tabular}{cccccll}
          \multirow{2}{*}{MLG}   &        \multicolumn{2}{c}{Centrosymmetric}    & \multicolumn{2}{c}{Noncentrosymmetric} & \multirow{2}{*}{Material}                                                 & \multirow{2}{*}{Tensor} \\
        &             MPG                    & $\boldsymbol\sigma$    & MPG               & $\boldsymbol\sigma$         &                                                                           &                         \\ \hline \hline
1 &  $\overline{1}$                       & arb. & 1                 & arb.      & \begin{tabular}[c]{@{}l@{}} Fe$_{\text{2}}$O$_{\text{3}}$   \end{tabular}                    & 
$\left(\begin{array}{ccc}{\sigma_{x x}} & {\sigma_{x y}} & {\sigma_{x z}} \\ {\sigma_{y x}} & {\sigma_{y y}} & {\sigma_{y z}} \\ {\sigma_{x z}} & {\sigma_{z y}} & {\sigma_{z z}}\end{array}\right) $             \\ \hline
     2   &             $2/m$   &    
                    \begin{tabular}[c]{@{}l@{}} $\parallel\textbf{a}_{\mathcal{C}_{2}}$ \\ $ \perp\mathcal{M}$ \end{tabular}  
                    &
\multicolumn{2}{c}{                    
                   \begin{tabular}{cc} $2$ & $\parallel\textbf{a}_{\mathcal{C}_{2}}$  \\ \hline $m$ & $ \perp\mathcal{M}$ \end{tabular}}    
 & \begin{tabular}[c]{@{}l@{}}BiCrO$_{\text{3}}$    \end{tabular} &  $\left(\begin{array}{ccc}{\sigma_{x x}} & {0} & {\sigma_{x z}} \\ {0} & {\sigma_{y y}} & {0} \\ {\sigma_{x z}} & {0} & {\sigma_{z z}}\end{array}\right)$      \                 \\ \hline \hline

$2'$ & $2'/m'$   &    
                    \begin{tabular}[c]{@{}l@{}} $\perp\textbf{a}_{\mathcal{TC}_{2}}$ \\ $ \in\mathcal{TM}$ \end{tabular}  
                    &
\multicolumn{2}{c}{                    
                   \begin{tabular}{cc} $2'$ & $\perp\textbf{a}_{\mathcal{TC}_{2}}$  \\ \hline $m'$ & $ \in\mathcal{TM}$ \end{tabular}}    
 & \begin{tabular}[c]{@{}l@{}}  CaMnO$_{\text{3}}$\cite{Vistoli2018}  \end{tabular} & $\left(\begin{array}{ccc}{\sigma_{x x}} & {\sigma_{x y}} & {\sigma_{x z}} \\ {-\sigma_{x y}} & {\sigma_{y y}} & {\sigma_{y z}} \\ {-\sigma_{x z}} & {-\sigma_{y z}} & {\sigma_{z z}}\end{array}\right)$      \            \\ \hline 
 
$2'2'2$ & $m'm'm$   &    
                    \begin{tabular}[c]{@{}l@{}} $\parallel\textbf{a}_{\mathcal{C}_{2}}$ \\ $ \perp\mathcal{M}_{z}$ \end{tabular}  
                    &
\multicolumn{2}{c}{                    
                   \begin{tabular}{cc} $2'2'2$ $m'm'2$ & $\parallel\textbf{a}_{\mathcal{C}_{2}}$  \\ \hline $m'm2'$ & $ \perp\mathcal{M}_{y}$ \end{tabular}}    
 & \begin{tabular}[c]{@{}l@{}}RuO$_{\text{2}}$\cite{Zhu2018} \\  CoNb$_{\text{3}}$S$_{\text{6}}$\cite{Ghimire2018}   \end{tabular} & $\left(\begin{array}{ccc}{\sigma_{x x}} & {\sigma_{x y}} & {0} \\ {-\sigma_{x y}} & {\sigma_{y y}} & {0} \\ {0} & {0} & {\sigma_{z z}}\end{array}\right)$      \             
                  
\end{tabular}
\caption{\textbf{Catalogue of Hall-vector admissible magnetic point groups in collinear antiferromagnets and selected material candidates.} First two rows list Type-I, and last two rows Type-III magnetic point groups (MPG), respectively. We list more material candidates and all magnetic symmetries allowing any Hall signal in SI Tab.~S2. If not referenced otherwise, the material candidate was obtained from MagnData database \cite{Gallego2016}. MLG marks magnetic Laue group.}
\end{table*}

\begin{acknowledgments}
We acknowledge discussions with M.~Kl\"{a}ui, M.~Jourdan, G.~Jakob, J.~\v{Z}elezn\'{y}, Y.~Mokrousov, and S.~Stemmer.
We acknowledge the use of the supercomputer Mogon at JGU (hpc.uni-mainz.de), the computing and storage facilities owned by parties and projects contributing to the National Grid Infrastructure MetaCentrum provided under the programme "Projects of Projects of Large Research, Development, and Innovations Infrastructures" (CESNET LM2015042) and support from the Alexander von Humboldt Foundation,  the Transregional Collaborative Research Center (SFB/TRR) 173 SPIN+X, EU FET Open RIA Grant no.~766566, the Grant Agency of the Charles University Grant no.~280815 and of the Czech Science Foundation Grant no.~14-37427. 
\end{acknowledgments}

\bibliographystyle{naturemag}

\end{document}